\documentclass[epj]{svjour}

\usepackage{amsmath}
\usepackage{amsfonts}

\DeclareMathSymbol{\R}{\mathalpha}{AMSb}{'122}

\begin{document}

\title{Thermalisation by a boson bath in a pure state}
\author{S. Camalet}
\institute{Laboratoire de Physique Th\'eorique de la Mati\`ere Condens\'ee, UMR 7600, 
Universit\'e Pierre et Marie Curie, Jussieu, Paris-75005, France}
\date{Received: date / Revised version: date }
\abstract{
We consider a quantum system weakly coupled to a large heat bath of harmonic oscillators. 
It is well known that such a boson bath initially 
at thermal equilibrium thermalises the system. We show that assuming a priori an equilibrium state 
is not necessary to obtain the thermalisation of the system.    
We determine the complete Schr\"odinger time evolution 
of the subsystem of interest for an initial pure product state of the composite system 
consisting of the considered system and the bath. 
We find that the system relaxes into canonical equilibrium for 
almost all initial pure bath states of macroscopically well-defined energy. 
The temperature of the system asymptotic thermal state is determined by the energy 
of the initial bath state as the corresponding microcanonical temperature. 
Moreover, the time evolution of the system is identical to the one obtained assuming 
that the boson bath is initially at thermal equilibrium at this temperature. 
A significant part of our approach is applicable to other baths and we identify the bath features 
which are requisite for the thermalisation.
 } 

\maketitle

\section{Introduction.}
\label{sec:intro}

The environment of a physical system plays an important role 
in its dynamics. The environmental degrees of freedom 
are responsible for the relaxation of the system into thermal equilibrium. 
As is well known, this irreversible evolution follows from 
Schr\"odinger dynamics for a system weakly coupled to a large heat 
bath, provided the bath is supposed to be initially at thermal equilibrium. 
However, this way to model the thermalisation process leaves open 
the question of how the system environment reaches 
its own equilibrium state. 
An equilibrium state is also assumed a priori in the most common derivation 
of the canonical equilibrium distribution. In this standard argumentation, 
the interaction between the system and its heat bath is neglected 
\cite{HI} 
and the microcanonical 
principle \cite{Diu} is invoked, i.e., the composite system consisting of the considered 
system and the bath is assumed to be in a microcanonical mixed state. 
The ensuing reduced state of the subsystem $S$ of interest is a canonical thermal mixture 
of its self-Hamiltonian eigenstates. Recently, it has been shown \cite{Tasaki,Mahler1,CT} 
that no statistical averaging is actually required to derive the canonical distribution for $S$. 
This thermal state is found for almost all {\it pure} states of the composite system 
of macroscopically well-defined energy. Canonical equilibrium is thus obtained as a direct 
consequence of the quantum mechanics principles. 

Nonetheless, to understand the thermalisation process, it is necessary to take into 
account the interaction between $S$ and the bath and to study the resulting 
Schr\"odinger dynamics of the isolated composite system. 
For the special case where the system-bath 
coupling Hamiltonian commutes with the self-Hamiltonian $H_S$ of $S$, 
it has been shown \cite{Zurek,Endo} that $S$ evolves into a statistical mixture 
of the eigenstates of $H_S$ for an initial ``generic'' pure 
product state of the composite system. 
But such a coupling cannot lead to a complete thermalisation of $S$. As 
the Hamiltonian $H_S$ is a constant of motion, the populations of the eigenstates of $H_S$ 
remain equal to their initial values. More recently, some results have been obtained  
regarding the thermalisation induced by a heat bath initially in a pure state 
for abstract system-bath interaction Hamiltonians defined in the eigenbasis of the 
non-interacting Hamiltonian. Analytical arguments showing that $S$ is at equilibrium 
most of the time are presented in \cite{Tasaki} and examples are studied numerically 
in \cite{Mahler2}.  

In this paper, we consider a system $S$ weakly coupled to a heat bath of harmonic oscillators. 
Numerous physical environments are well modelled by such 
a boson bath which is thus often used to describe dissipation in quantum systems \cite{QDS}. 
From this perspective, the bath is usually assumed at thermal equilibrium at initial time. 
Here we study the reduced dynamics of $S$ for an initial {\it pure} product state of the composite 
system  consisting of $S$ and its bosonic environment. 
Using the resolvent operator formalism, we derive this dynamics from the Schr\"odinger 
equation of the isolated composite system. Since we are interested 
in the thermalisation process, we suppose $S$ is initially in a pure state.
For the bath, we consider a ``typical" state of macroscopically 
well-defined energy. Whereas our results are obtained for a boson bath, 
a significant part of the formalism developed is applicable to other environments. 
The paper is organised as follows. 
In the following section, we present the model and the initial bath state. 
The reduced dynamics of the subsystem $S$ is derived in Sect.~\ref{sec:rd}. 
We then determine the asymptotic state of $S$ and the relaxation 
to this state in the weak coupling regime in Sect.~\ref{sec:wcr}.       
The results and their potential validity for other environments 
are discussed in Sect.~\ref{sec:discussion}.

\section{Model}
\label{sec:model}

We write the total Hamiltonian of the composite system consisting 
of the system $S$ of interest and its bosonic environment as 
\begin{equation}
H=H_S+H_B+H_I \label{H}
\end{equation}
where $H_S$ and $H_I$ describe, respectively, the intrinsic dynamics of $S$ 
and its coupling to the boson bath. The diagonalised bath Hamiltonian $H_B$ reads
\begin{equation}
H_B=\sum_q \omega_q b^\dag_q b^{\phantom{\dag}}_q \label{HB}
\end{equation}
where $b^\dag_q$ ($b_q$) are bosonic creation (annihilation) operators and 
the sum runs over the $N$ harmonic modes of the bath. The eigenfrequencies $\omega_q$ 
depend on the specific bath considered \cite{QDS}. The thermodynamic limit 
$N \gg 1$ is assumed throughout this paper. 
In the following, the eigenstates of $H_S$ and $H_B$ are denoted by Latin and Greek letters, 
respectively, i.e., 
\begin{eqnarray}
H_S |k\rangle &=& \epsilon_k |k\rangle \nonumber \\
H_B |\alpha \rangle &=& E_\alpha |\alpha \rangle 
\end{eqnarray} 
where $\epsilon_k$ and $E_\alpha$ are the respective eigenvalues. 
For the system $S$, we consider a discrete spectrum of nondegenerate 
eigenergies $\epsilon_k$. A system with continuous spectrum cannot 
relax into thermal equilibrium. For example, in the case of a free damped particle, 
the width of an initially localised state grows indefinitely. The bath influences 
significantly this diffusion process \cite{Gert}. For the bath, an eigenstate $|\alpha \rangle$ 
corresponds to a set of harmonic mode occupation numbers $\{ n_q \}$ 
and $E_\alpha=\sum_q \omega_q n_q$. In the thermodynamic limit, the eigenenergy 
spectrum of the bath 
can be characterised by a density of states $n (E)$ given by the standard following definition 
\cite{Diu}. For a macroscopic energy $E \sim N$ and an energy $\delta E \ll E$ but far larger 
than the maximum level spacing of $H_B$, 
the number $D$ of states $|\alpha \rangle$ 
such that $E<E_\alpha<E+\delta E$, is practically proportionnal to $\delta E$ and 
$n (E)=D/\delta E$. Moreover, this density satisfies the Boltzmann's relation
\begin{equation}
\ln \left[ n (E) \delta E \right]  \simeq N s(E/N)  \label{n}
\end{equation}  
where $s$ is the bath entropy per oscillator. We use the units $\hbar=k_B=1$ throughout this paper. 
As the bath entropy $Ns$ and energy $E$ are extensive variables, the entropy per oscillator 
$s$ assumes finite values for finite energies per oscillator $E/N$.    

The interaction Hamiltonian $H_I$ 
depends on the degrees of freedom of both the bath and the system $S$. 
It is useful for the following to expand it as 
\begin{equation}
H_I = \sum_{k,l} |k\rangle\langle l | U_{kl} \label{HI}
\end{equation}
where $U_{kl}$ are bath operators which obey $U^\dag_{kl}=U^{\phantom{\dag}}_{lk}$. 
This form is generally valid for a composite system. 
We consider here the interaction operators 
\begin{equation}
U_{kl} =  \sum_q \kappa_q^{kl} ( b^\dag_q + b^{\phantom{\dag}}_q) . \label{U}
\end{equation}
Numerous environments can be 
modeled by such a linear coupling to the positions of independent harmonic oscillators \cite{QDS}. 
The influence on $S$ of the bosonic environment described by $H_B$ and the coupling operators 
\eqref{U} can be characterised by the standard spectral densities
\begin{equation}
J_{ij,kl} (\omega) = \sum_q \kappa_q^{ij} \kappa_q^{kl}  \delta(\omega-\omega_q) \label{sd}
\end{equation}
which are finite continuous functions of $\omega >0$ in the thermodynamic limit. Since 
$(\kappa_q^{ji})^*=\kappa_q^{ij}$, they satisfy $J_{kl,ij}=J^*_{ji,lk}=J_{ij,kl}$.

In this paper, we study the Schr\"odinger time evolution of the system $S$ following 
from an initial pure product state of the composite system. We assume that the initial bath state 
\begin{equation}
|\psi \rangle = \sum_{|\alpha\rangle \in {\cal H}_E} \psi_\alpha |\alpha \rangle \label{psi}
\end{equation} 
is a normalised state of the Hilbert space ${\cal H}_E$ defined by the 
$D$ states $|\alpha \rangle$ 
such that $E<E_\alpha<E+\delta E$. 
In the following, we show that almost all normalised states in ${\cal H}_E$ 
lead to the same behavior of $S$. To do so, we use the uniform measure $\mu$ 
on the unit sphere in ${\cal H}_E$ given by   
\begin{equation}
\mu \big( \{\psi_\alpha\} \big) = \frac{(D-1)!}{\pi^D} \delta 
\Big( 1-\sum_{|\alpha\rangle \in {\cal H}_E} |\psi_\alpha|^2 \Big). \label{mu}
\end{equation}
With this normalised measure, we will prove that a subset of bath states \eqref{psi} resulting 
in the same time evolution of $S$, is of size unity in the thermodynamic limit. 
In \cite{CT}, the uniform measure is discussed for composite system states $\Phi$ 
of macroscopically well-defined energy. The product states considered here and the corresponding 
measure \eqref{mu} can 
be obtained by performing ideal quantum measurements of the observable $H_S$ on the states $\Phi$. 
We remark that other choices are possible to describe an initial bath state 
of macroscopically well-defined energy. 
For example, the space ${\cal H}_E$ can be replaced in \eqref{psi} by the Hilbert space 
${\cal H}'_E$ defined by the eigenstates $|\alpha \rangle$ such that $E_\alpha<E$. 
In this case, the appropriate measure 
is given by the expression \eqref{mu} with the replacements ${\cal H}_E \rightarrow {\cal H}'_E$ 
and $D \rightarrow D'$ where $D'=\int_0^E n(E') dE'$ is the dimension of ${\cal H}'_E$. 
Due to the exponential $N$-dependence of the density of states $n$, 
the number of states $|\alpha \rangle$ in ${\cal H}'_E$ but not in ${\cal H}_{E-\delta E}$ is negligible
and hence the energy of a typical state in ${\cal H}'_E$ is practically equal to $E \sim N$. 
For the same reason, the results presented in the remainder of the paper are valid for such a state.

\section{Reduced dynamics}
\label{sec:rd}

We derive here the reduced dynamics of the subsystem $S$ for an initial disentagled state
\begin{equation}
\Omega = \rho_S \rho_B \label{Omega}
\end{equation} 
where $\rho_S$ and $\rho_B$ are density matrices of $S$  and the bath, respectively. 
The initial bath state can be, for example, a thermal mixed state, 
$\rho_B \propto \exp(-H_B/T)$, or a pure state, $\rho_B=|\psi\rangle\langle \psi |$. 
In this paper, we are concerned with the thermalisation induced by the pure states 
given by  \eqref{psi}. However, the calculation presented in this section is not restricted 
to pure initial bath states and it is interesting for the following to consider also thermal states. 
To obtain the time evolution of $S$, we first  write its reduced density matrix at time 
$t > 0$ as 
\begin{equation}
\rho(t) = \frac{i}{2\pi} \int_{\R+i\eta} dz  e^{-izt}  {\tilde \rho} (z) \label{rhot} 
\end{equation}  
where $\eta$ is a positive real number. The Laplace transform ${\tilde \rho} (z)$ of the state $\rho(t)$ 
is given by
\begin{equation}
{\tilde \rho} (z) = \mathrm{Tr}_B \left[  {\cal G} (z) \Omega \right] \label{rhoz} 
\end{equation}  
where $\mathrm{Tr}_B$ denotes the partial trace over the bath 
degrees of freedom and ${\cal G}(z)=\left( z-{\cal L} \right)^{-1}$ is the resolvent of 
the Liouvillian ${\cal L}$ defined as ${\cal L} A=[H,A]$ for any operator $A$. The Liouvillians 
${\cal L}_S$, ${\cal L}_B$ and ${\cal L}_I$ corresponding to the Hamiltonians $H_S$, $H_B$ 
and $H_I$, respectively, are given by similar definitions. An equation for ${\tilde \rho} (z)$ can 
be derived from the Schr\"odinger time evolution of the composite system consisting of $S$ 
and its environment using the projection superoperators ${\cal P}$ and ${\cal Q}$ 
defined by ${\cal Q}=1-{\cal P}$ and ${\cal P} A=\mathrm{Tr}_B (A) \rho_B $ where $A$ is any 
operator. With these superoperators, we deduce from $( z-{\cal L}){\cal G}(z)\Omega=\Omega$ 
the coupled equations \cite{ROM1,RSB}
\begin{eqnarray} 
{\cal P}(z-{\cal L}){\cal P}{\cal P}{\cal G}(z) \Omega + {\cal P}(z-{\cal L}){\cal Q} {\cal Q}{\cal G}(z) \Omega&=&
\Omega \nonumber \\
{\cal Q}(z-{\cal L}){\cal P}{\cal P}{\cal G}(z)\Omega + {\cal Q}(z-{\cal L}){\cal Q}{\cal Q}{\cal G}(z)\Omega&=&0
\end{eqnarray}
for the operators ${\cal P}{\cal G}(z) \Omega$ and ${\cal Q}{\cal G}(z) \Omega$. Solving the latter for 
${\cal Q}{\cal G}(z) \Omega$ in terms of ${\cal P}{\cal G}(z) \Omega$ and then substituting in the former yields 
\begin{equation}
\mathrm{Tr}_B \left[ \left( z-{\cal L} - {\cal L} {\cal Q} (z-{\cal Q}{\cal L}{\cal Q})^{-1} {\cal Q} {\cal L} \right) 
\rho_B {\tilde \rho} (z) \right] = \rho_S . \label{rd}
\end{equation}  
This equation describes the complete time evolution of $S$. 
Since we are interested in the limit of weak coupling between $S$ and the bath, 
we expand $(z-{\cal Q}{\cal L}{\cal Q})^{-1}$ in the Liouvillian ${\cal L}_I$.
With this expansion and the properties $\mathrm{Tr}_B ({\cal L}_B A)=0$ and 
$\mathrm{Tr}_B ({\cal L}_S A)={\cal L}_S \mathrm{Tr}_B(A)$ where $A$ is any operator, 
we find 
\begin{eqnarray}
\rho_S &=& z {\tilde \rho} (z) - [H_S,{\tilde \rho} (z)] 
- \mathrm{Tr}_B \left[ {\cal L}_I \rho_B {\tilde \rho} (z) \right] \label{rd2} \\
&&- \mathrm{Tr}_B \bigg[ \sum_{n > 0} \Big( 
{\cal L}_{I} {\cal Q}  \frac{1}{z-{\cal L}_{SB}} {\cal Q} \Big)^n 
({\cal L}_{B}+{\cal L}_{I}) \rho_B {\tilde \rho} (z) \bigg]  \nonumber 
\end{eqnarray}  
where ${\cal L}_{SB}={\cal L}_S+{\cal L}_B$. 

It is now convenient to expand the Laplace transform ${\tilde \rho} (z)$ in the basis 
$|k\rangle\langle l|$ of the Liouville space of $S$ as 
\begin{equation}
{\tilde \rho} (z) = \sum_{k,l} |k\rangle\langle l| r_{kl} (z) \label{rkl}
\end{equation}
where $r_{kl} (z)$ are scalar functions of $z$. These components are solutions of the coupled equations 
\begin{equation}
(z-\omega_{ij}) r_{ij} (z) - \sum_{k,l} \Sigma_{ij}^{kl} (z) r_{kl} (z) =  \langle i| \rho_S |j\rangle  \label{eqr}
\end{equation}
where $\omega_{ij}=\epsilon_i-\epsilon_j$. The influence of the environment is described by the self-energies 
\begin{eqnarray}
\Sigma_{ij}^{kl} (z) &=& \big\langle i\big| \mathrm{Tr}_B \bigg[ 
\sum_{n \ge 0} \Big( {\cal L}_{I} {\cal Q} \frac{1}{z-{\cal L}_{SB}}  {\cal Q}  \Big)^n {\cal L}_I  \nonumber \\
& &\qquad \qquad \quad  \times 
  \frac{z-\omega_{kl}}{z-{\cal L}_{SB}} \rho_B \big|k\big\rangle \big\langle l \big| \bigg] 
  \big|j \big\rangle. \label{Sigma}
\end{eqnarray}
To derive this expression from \eqref{rd2} we have grouped the terms of same order in $H_I$ 
using $1+{\cal B}^{-1}({\cal A}-{\cal B})={\cal B}^{-1} {\cal A}$ with ${\cal A}=z-{\cal L}_{S}$ 
and ${\cal B}=z-{\cal L}_{SB}$. For the usually considered thermal bath state 
$\rho_B \propto \exp(-H_B/T)$, the above derivation is simpler as ${\cal L}_{B}\rho_B=0$ and hence 
${\cal Q}$ and ${\cal L}_{SB}$ commute. 
In this case, the superoperator $(z-\omega_{kl})/(z-{\cal L}_{SB})$ in \eqref{Sigma} 
can be replaced by $1$ \cite{ROM}. 
We remark that, as $\mathrm{Tr}({\cal L}_I A)=0$ for any operator $A$, 
the self-energies \eqref{Sigma} satisfy the relations $\sum_i \Sigma_{ii}^{kl} (z)=0$ 
which ensure that the trace of the density matrix $\rho(t)$ is conserved. 
To obtain the reduced dynamics \eqref{eqr} 
we have only used the form \eqref{H} of the total Hamiltonian describing 
the system $S$, the bath and their coupling. The bosonic nature of the bath and the specific 
coupling bath operators	 \eqref{U} do not play any role in the above calculation. 

The equations \eqref{eqr} can be solved in $r_{kl}(z)$ by writing these components as power series 
in $H_I$. However, the perturbative solutions obtained by this method give the state $\rho(t)$ 
of $S$, through \eqref{rkl} and \eqref{rhot}, only for short times $t$. 
To determine the long time evolution of $S$ for weak coupling to the bath, 
we proceed as follows. The coupled equations \eqref{eqr} can be written as 
${\bf M}(z) {\bf r}(z) = {\bf r}_S$ 
where ${\bf r}(z)$ and ${\bf r}_S$ are vectors of components $r_{kl} (z)$ and 
$\langle k| \rho_S |l\rangle$, respectively, and ${\bf M}(z)$ is a matrix of elements 
$M_{ij,kl}(z)=(z-\omega_{ij})\delta_{ki}\delta_{lj}-\Sigma_{ij}^{kl} (z)$. 
This matrix equation can be formally solved by inverting ${\bf M}(z)$ with the help of 
its eigenvalues $\lambda_n(z)$ and eigenvectors ${\bf u}_n(z)$ and the eigenvectors 
${\bf v}_n(z)$ of the 
transpose ${\bf M}(z)^T$. With these definitions, we write the solution of \eqref{eqr} as 
\begin{equation}
{\bf r}(z) = \sum_n \frac{1}{\lambda_n(z)} {\bf u}_n(z) {\bf v}_n(z)^T {\bf r}_S . \label{M-1}
\end{equation}
In absence of coupling to the bath, i.e., for $H_I=0$, the matrix ${\bf M}(z)$ is diagonal and 
$\{ \lambda_n(z) \}=\{ z-\omega_{kl}\}$. As a result, the populations $\langle k| \rho(t) |k\rangle$ 
are constant and the coherence $\langle k| \rho(t) |l\rangle$ oscillates with the frequency $\omega_{kl}$. 
For a weak coupling, the eigenvalues $\lambda_n(z)$ can be determined as power series 
expansions in $H_I$ as discussed in the next section. 
As a consequence of the property $\sum_i \Sigma_{ii}^{kl} (z)=0$, $\lambda_0(z)=z$ remains 
an eigenvalue of ${\bf M}(z)$ for any coupling strength. Moreover, the components 
of the corresponding eigenvector ${\bf v}_0(z)$ are $\delta_{ij}$ for any $z$. 
The term $n=0$ of \eqref{M-1} thus simplifies to 
${\bf u}_0(z)/z$. This term results in a steady contribution to the state $\rho(t)$ of $S$. 
For the model considered in this paper, any eigenvalue of index $n \ne 0$ 
manifests a branch cut on the real axis and $\lambda_n(\omega+i0^+)$ is a nonvanishing 
function of the real argument $\omega$. 
Consequently, the terms $n \ne 0$ of \eqref{M-1} correspond to contributions to $\rho(t)$ 
which vanish at asymptotic times. 
In conclusion, the components $r^{\infty}_{kl}$ of the asymptotic state $\rho_\infty$ of $S$ 
are that of ${\bf u}_0(i0^+)$, i.e., they are given by
\begin{equation}
\omega_{ij} r^\infty_{ij} + \sum_{k,l} \Sigma_{ij}^{kl} (i0^+) r_{kl}^\infty =  0 . \label{eqasyr}
\end{equation}
Using this equation and the expressions \eqref{Sigma}, $\rho_\infty$ can be determined for weak 
coupling to the bath by writing the components $r^{\infty}_{kl}$ as power series in $H_I$.

\section{Weak coupling regime}
\label{sec:wcr}

In this section, we determine the asymptotic state $\rho_\infty$ of the system $S$ 
for weak coupling to the boson bath and a typical initial bath state \eqref{psi} of 
macroscopic energy $E$. We also discuss 
the terms $n \ne 0$ of the expression \eqref{M-1} which describe the decoherence 
and relaxation of $S$ towards $\rho_\infty$. More precisely, 
we evaluate the lowest relevant orders in $H_I$ of the state $\rho_\infty$ and 
eigenvalues $\lambda_n(z)$. To do so, we first derive tractable power series expansions of 
the self-energies \eqref{Sigma} as follows. By writing $(z-{\cal L}_{SB})^{-1}$ as the Laplace 
transform of $\exp(-it{\cal L}_{SB})$ and ${\cal L}_I$ in terms of the bath operators $U_{kl}$ 
using \eqref{HI}, we express $\Sigma_{ij}^{kl} (z)$ in terms of the Heisenberg 
operators $U_{kl} (t)=\exp(itH_B) U_{kl} \exp(-itH_B)$. We will see that, 
for weak coupling to the bath, the reduced dynamics of the subsystem $S$ 
is essentially determined by the averages $ \langle U_{kl} (t) \rangle$ and 
$ \langle U_{ij} (t)U_{kl} (t+\tau)  \rangle$ where 
$\langle A \rangle=\langle \psi | A | \psi \rangle$ for any bath operator $A$. 
As shown below, explicit expressions can be obtained for these expectation values.

\subsection{Correlations of the bath interaction operators}
\label{sec:tdc}

We show here that, in the thermodynamic limit, almost all bath states \eqref{psi} 
lead to the same averages $ \langle U_{kl} (t) \rangle$ and 
$ \langle U_{ij} (t)U_{kl} (t+\tau)  \rangle$. We will obtain this result by computing the 
Hilbert space averages \cite{Mahler1} and variances of these expectation values 
resulting from the measure \eqref{mu}. 
For example, the Hilbert space average of $ \langle U_{kl} (t) \rangle$ reads 
\begin{equation}
\overline{\langle U_{kl} (t) \rangle} = 
\prod_{\alpha} \int  d^2\psi_\alpha \mu (\{\psi_\alpha\})  
\sum_{\alpha,\beta} \psi^*_\alpha \psi_\beta 
\langle \alpha | U_{kl} (t) |\beta \rangle 
\end{equation}
where $d^2\psi_\alpha=d\mathrm{Re}\psi_\alpha d\mathrm{Im}\psi_\alpha$ and 
the product and the sums run over the bath eigenstates $|\alpha \rangle \in {\cal H}_E$. 
To evaluate this integral, we first note that, for a bath size $N \gg 1$, 
\begin{equation}
\prod_{\alpha \ne \alpha(p)} \int  d^2\psi_\alpha \mu (\{\psi_\alpha\})  
\simeq \prod_{p=1}^{n} \frac{D}{\pi}e^{-D | \psi_{\alpha(p)} |^2} \label{reddis}
\end{equation}
where the $n$ components $\psi_{\alpha(p)}$ are fixed \cite{QSC}.
This reduced distribution and the expansion \eqref{psi} then yield 
$\overline{ \langle U_{kl} (t) \rangle}=\langle  U_{kl} \rangle_{E}$ where 
\begin{equation}
\langle  A \rangle_{E} = \frac{1}{D} 
\sum_{|\alpha \rangle \in {\cal H}_E} \langle \alpha | A |\alpha \rangle 
\label{ma}
\end{equation}  
denotes the microcanonical average of any bath operator $A$. 
We define the variance $\sigma_1^2$ as the Hilbert space average of 
$|\langle U_{kl} (t) \rangle - \langle  U_{kl} \rangle_{E} |^2$ and find 
\begin{equation}
\sigma^2_1=\frac{1}{D^2}\sum_{|\alpha\rangle,|\beta\rangle \in {\cal H}_E} 
|\langle \alpha | U_{kl}|\beta \rangle |^2 
< \frac{1}{D} \langle U^{\phantom{\dag}}_{kl}  U^\dag_{kl} \rangle_E .
\end{equation} 
The upper bound is simply obtained by replacing the sum 
over the states $|\beta\rangle \in {\cal H}_E$ by a sum over all the bath eigenstates $|\beta\rangle$. 
For $ \langle U_{ij} (t)U_{kl} (t+\tau)  \rangle$, we find the Hilbert space average 
$ \langle U_{ij} U_{kl} (\tau)   \rangle_E$ and variance 
\begin{eqnarray}
\sigma^2_2(\tau)&=&\frac{1}{D^2} 
\sum_{|\alpha\rangle,|\beta\rangle \in {\cal H}_E} 
|\langle \alpha | U_{ij} U_{kl} (\tau) |\beta \rangle |^2 \nonumber \\
&<&  \frac{1}{D} \langle U^{\phantom{\dag}}_{ij} U^{\phantom{\dag}}_{kl}(\tau)U^{\dag}_{kl} (\tau) 
U^{\dag}_{ij}  \rangle_E \label{sigma2} . 
\end{eqnarray} 

We now determine the microcanonical correlation functions $ \langle U_{ij} U_{kl} (\tau)  \rangle_E$ 
using the equality of the occupation number $ \langle b^\dag_q b^{\phantom{\dag}}_q \rangle_E$ 
to the canonical average $ \langle b^\dag_q b^{\phantom{\dag}}_q \rangle_T$ 
with the appropriate temperature $T$. 
The canonical average of any bath operator $A$ for temperature $T$ reads
\begin{equation}
\langle  A \rangle_{T} = \frac{1}{Z} 
\sum_{\alpha} e^{-E_\alpha/T} \langle \alpha | A |\alpha \rangle 
\end{equation}  
where $Z=\sum_{\alpha} \exp(-E_\alpha/T)$. It can be established, by the following 
standard arguments \cite{Diu}, 
that $ \langle b^\dag_q b^{\phantom{\dag}}_q \rangle_E=\langle b^\dag_q b^{\phantom{\dag}}_q \rangle_T$  
for $T$ equals to the bath microcanonical temperature corresponding to the energy $E$, i.e.,
\begin{equation}
\frac{1}{T} = s' \left( \frac{E}{N} \right)   \label{T}
\end{equation}  
where $s'$ is the derivative of the bath entropy per oscillator $s$ with respect to the energy $E/N$. 
The microcanonical distribution of the harmonic mode occupation $n_q$ 
is $P(n_q)={\hat n}(E-n_q \omega_q )/n(E)$ where ${\hat n}$ is the density of states 
of the system consisting of the $N-1$ other bath modes. 
This density satisfies the Boltzmann's relation \eqref{n} with the corresponding entropy ${\hat s}$. 
Expanding this entropy in the energy $n_q \omega_q$ and taking into account 
that ${\hat s}=s$ in the thermodynamic limit, results in $P(n_q) \propto \exp(-n_q\omega_q /T)$ 
where $T$ is given by \eqref{T}. Using this distribution, we obtain 
\begin{equation}
\begin{array}{l}
\displaystyle  \  \langle U_{ij} U_{kl}  (\tau) \rangle_E = \langle U_{ij} U_{kl}  (\tau) \rangle_T  \label{UtauU} \\ \\
 \displaystyle \   \qquad = \int_0^\infty d\omega J_{ij,kl}(\omega) 
 \left[ \frac{\cos ( \omega \tau)}{\tanh ( \omega/2T)} + i \sin ( \omega \tau) \right]  . 
\end{array}
\end{equation}       

\begin{sloppypar}
It follows from this result that $\sigma^2_1=O( \exp(-N))$. 
Consequently, the overwhelming majority of bath states \eqref{psi} satisfy 
\begin{equation}
\langle U_{kl} (t) \rangle=0. \label{Uklt} 
\end{equation}
The dynamical fluctuations of $\langle U_{kl} (t) \rangle$ around zero 
disappear in the thermodynamic limit  for typical bath states $|\psi \rangle$. 
By generalising the above derivation to the microcanonical distribution 
of two harmonic mode occupation numbers, it can be shown that the left side of the inequality 
\eqref{sigma2} is equal to a canonical average. Moreover, as the operators $U_{kl}$ 
are linear combinations of the operators $b^\dag_q$ and $b_q$, the upper bound of 
$\sigma^2_2 (\tau)$ can be rewritten as   
$(\langle U_{ij}  U_{ji} \rangle_T \langle U_{kl}  U_{lk} \rangle_T 
+|\langle U_{ij}  U_{kl}(\tau) \rangle_T|^2
+ |\langle U_{ij}  U_{lk}(\tau) \rangle_T|^2 )/D$. 
As a result, the variance $\sigma^2_2(\tau)$ also vanishes exponentially in the limit $N \gg 1$ 
and hence, using \eqref{UtauU},  
\begin{equation} 
\langle U_{ij} (t)U_{kl} (t+\tau)  \rangle =\langle U_{ij}U_{kl}(\tau)  \rangle_T  . \label{UtUt'}
\end{equation}
\end{sloppypar}

\subsection{Asymptotic state, decoherence and relaxation}
\label{sec:asr}

We deduce from the result \eqref{Uklt} that the first order contribution in $H_I$ to the self-energy 
\eqref{Sigma} with $\rho_B=|\psi\rangle \langle \psi |$ vanishes. 
To determine the second order contribution, we first rewrite it as a sum of four terms 
using ${\cal Q}A=A-\rho_B \mathrm{Tr}_B (A)$ where A is any operator. 
As the first-order self-energies vanish and 
$\mathrm{Tr}_B [(z-{\cal L}_{SB})^{-1} \rho_B \mathrm{Tr}_B(A)]
=\mathrm{Tr}_B[(z-{\cal L}_{SB})^{-1}A]$ where $A$ is any operator, only one 
term remains and 
\begin{equation}
\Sigma_{ij}^{kl} (z)=(z-\omega_{kl})
\big\langle i \big| \mathrm{Tr}_B \bigg[ \Big( {\cal L}_I \frac{1}{z-{\cal L}_{SB}} \Big)^2 \label{Sigma21}
\rho_B \big| k \big\rangle \big\langle l \big| \bigg] \big|j \big\rangle
\end{equation}   
to second order in $H_I$, where $\rho_B=|\psi \rangle \langle \psi |$.
A lengthy but straightforward calculation then gives 
\begin{eqnarray}
\Sigma_{ij}^{kl} (z) &=& -i \int_0^\infty dt e^{itz} 
\Big[ \delta_{ik} \sum_p e^{it\omega_{pk}} \langle U_{lp} U_{pj}(t) \rangle_T \nonumber \\
& &  \qquad  \qquad  +\delta_{jl} \sum_p e^{it\omega_{lp}} \langle U_{ip} (t) U_{pk} \rangle_T   \label{Sigma2} \\
& & - e^{it\omega_{li}} \langle U_{lj} (t) U_{ik} \rangle_T 
- e^{it\omega_{jk}} \langle U_{lj} U_{ik}(t) \rangle_T   \Big]  \nonumber
\end{eqnarray} 
to second order in $H_I$. The final expression of the calculation 
has been rewritten with the help of the equality \eqref{UtUt'}. 
The temperature $T$, given by \eqref{T}, is the microcanonical temperature 
of the bath corresponding to the energy $E$ of its initial state. 

Since $\omega_{ij} \ne 0$ for $i \ne j$ and the self-energies $\Sigma_{ij}^{kl}$ are 
at least of second order, 
\eqref{eqasyr} leads to $r^{\infty}_{kl}=P_k \delta_{kl} + O(H_I^2)$. 
Writing this equation to second order for $i=j$ and using \eqref{Sigma2}, we find 
\begin{equation}
\sum_k \Gamma_{ik} P_k - \sum_k \Gamma_{ki} P_i = 0 . \label{me}
\end{equation}   
The zeroth-order populations $P_i$ are thus the steady solutions of a master equation. 
From \eqref{Sigma2} and \eqref{UtauU}, we deduce the (positive) rates
\begin{equation}
\Gamma_{ik} = \frac{K_{ik}(\epsilon_i-\epsilon_k)}{e^{(\epsilon_i-\epsilon_k)/T}-1} \label{Gamma}
\end{equation}   
where $K_{ik}(\omega)=2\pi [\Theta(\omega)J_{ik,ki}(\omega)- \Theta(-\omega)J_{ik,ki}(-\omega)]$. 
As $J_{ki,ik}=J_{ik,ki}$, the rates \eqref{Gamma} satisfy the detailed balance relation, 
\begin{equation}
\Gamma_{ki} e^{-\epsilon_i/T}=\Gamma_{ik} e^{-\epsilon_k/T}.
\end{equation} 
Consequently, assuming the matrix ${\bf G}$ of elements 
$G_{ik}=-\Gamma_{ik}+\delta_{ik}\Gamma_{ki}$ is irreducible \cite{Diu}, 
the zeroth-order asymptotic state of $S$ is the canonical thermal state, 
\begin{equation}
\rho_\infty = \frac{e^{-H_S/T}}{\mathrm{Tr} (e^{-H_S/T})}. \label{therm}
\end{equation}    
The following special case is worth mentioning. 
If the self-Hamiltonian $H_S$ of $S$ and the interaction Hamiltonian $H_I$ commute, i.e., $U_{kl}=0$ 
for $k \ne l$, the populations $\langle k | \rho(t) | k \rangle$ remain equal to their respective 
initial values. In this case, the only effect of the environment is to destroy the coherences 
$\langle k | \rho(t) | l \rangle$ \cite{Zurek,Endo}.  

We now discuss the terms $n \ne 0$ of the expression \eqref{M-1} which describe the decoherence 
and relaxation of $S$ towards the thermal equilibrium state \eqref{therm}. 
We first consider a frequency $\omega_{kl} \ne 0$ such that $\omega_{ij} \ne \omega_{kl}$ 
for any $(i,j) \ne (k,l)$. For such a frequency, 
we find $\lambda_n (z)=z-\omega_{kl}-\Sigma_{kl}^{kl}(z)$ to second order in $H_I$. 
The corresponding term in \eqref{M-1} describes the time evolution of the coherence 
$\langle k | \rho(t) | l \rangle$. The influence of the environment on this coherence is easier to 
understand for $z=\omega+i0^+$ where $\omega$ is real \cite{ROM1,RSB}. 
For weak coupling to the bath, 
the perturbative corrections to $\lambda_n (\omega+i0^+)$ can be neglected except for 
$\omega \simeq \omega_{kl}$. From \eqref{Sigma2} and \eqref{UtauU}, we obtain
\begin{equation}
\begin{array}{l}
\displaystyle \Sigma_{kl}^{kl}(\omega_{kl}+i0^+) = -\frac{i}{2} \sum_{p \ne k,l} 
\left( \Gamma_{pk} + \Gamma_{pl} \right)-i S_{kl}  \label{do} \\
\displaystyle \quad + \frac{1}{2\pi} \sum_p \wp \int
d\omega \frac{1}{e^{\omega/T}-1} \left( 
\frac{K_{pk}(\omega)}{\omega-\omega_{pk}} -  \frac{K_{pl}(\omega)}{\omega-\omega_{pl}} \right)  .  
\end{array}
\end{equation}      
to second order in $H_I$, where $\wp$ denotes the Cauchy principal value and 
$S_{kl}=\pi \sum_q (\kappa_q^{kk}-\kappa_q^{ll})^2 \coth(\omega_q/2T) \delta(\omega_q)$. 
The negative imaginary part, in 
the first line, leads to an exponential decay of $\langle k | \rho(t) | l \rangle$ at long times 
\cite{QDS,ROM1,RSB}. The real part results in a slight shift of the oscillation frequency 
of this coherence. As is well known, there exists a short-time regime which is not 
accurately described by the approximation 
$\lambda_n(\omega+i0^+) \simeq \omega -\omega_{kl}-\Sigma_{kl}^{kl}(\omega_{kl}+i0^+)$. 
However, the behavior of $S$ at short times $t$ is also determined 
by the second-order self-energies \eqref{Sigma2} as $\rho(t)$ is well approximated 
by an expansion to second order derived from \eqref{eqr}. 
If $\omega_{kl}={\bar \omega} \ne 0$ for several $(k,l)$, 
the second-order contributions to the eigenvalues $\lambda_n(z)=z-{\bar \omega}+\cdots$ 
are obtained by diagonalising a matrix of elements $\Sigma_{ij}^{kl}({\bar \omega}+i0^+)$ (to second order) 
where $(i,j)$ and $(k,l)$ are such that $\omega_{ij}=\omega_{kl}={\bar \omega}$. 
The corresponding coherences of $\rho(t)$ vanish at asymptotic times. 
For the eigenvalues ${\lambda}_n (z)=z+\cdots$, it can be shown 
that $\lambda_n (i0^+)=i\gamma_n$ to second order,  
where $\gamma_n$ are the positive real eigenvalues of the matrix ${\bf G}$. 
One of them vanishes and corresponds to the asymptotic state 
of $S$ given by \eqref{me}. The other ones are the characteristic rates of 
the long-time relaxation of the populations $\langle k | \rho(t) | k \rangle$. 
In conclusion, for weak coupling to the bath, the complete time evolution of $S$ 
is determined by the second-order 
self-energies \eqref{Sigma2} and hence is identical to the one obtained in the usual 
case of a bath initially at thermal equilibrium, $\rho_B = \exp(-H_B/T)/Z$. 

\section{Discussion of the results}
\label{sec:discussion}

We discuss here the relative importances of the different environment features in the obtention 
of the asymptotic thermal state \eqref{therm}. We first reiterate that the derivation 
of the reduced dynamics of the subsystem $S$, presented in Sect.~\ref{sec:rd}, relies 
essentially on the form \eqref{H} of the total Hamiltonian of the composite system. 
To obtain then from \eqref{eqasyr} 
the equations \eqref{me} for the zeroth-order asymptotic populations, 
the vanishing of the first order contributions to the self-energies \eqref{Sigma} is not necessary. 
It can be proved from the expression \eqref{Sigma} that the equality 
\eqref{Sigma21} is correct to second order in $H_I$ in the limit $z \rightarrow i0^+$ 
for $i=j$ and $k=l$, for any Hamiltonian \eqref{H}. Moreover, for a pure bath state, 
$\rho_B = |\psi \rangle \langle \psi |$, the right side of \eqref{Sigma21} for $i=j$ and $k=l$ 
can be written as $z\Upsilon_{ik}(z)-\delta_{ik} z \sum_p \Upsilon_{pi}(z)$ without any further 
assumption.  The solution of the ensuing equation \eqref{me} is the canonical distribution 
if the rates $\Gamma_{ik}$ satisfy the detailed balance relation. 
This relation results directly from \eqref{Sigma2} as this second-order self-energy leads to 
the Fermi golden rule like expression
\begin{equation}
\Gamma_{ik} = \frac{2\pi}{Z} \sum_{\alpha,\beta} e^{-E_\alpha/T} 
\delta(E_\beta+\epsilon_i-E_\alpha-\epsilon_k) 
|\langle \beta |U_{ik}| \alpha \rangle|^2 \label{Gammacan} 
\end{equation}     
which obviously obeys $\Gamma_{ki}\exp[(\epsilon_k-\epsilon_i)/T]=\Gamma_{ik}$. 
The specific form \eqref{UtauU} of the correlations of the bath interaction operators $U_{kl}$ 
is then not decisive to obtain the detailed balance relation.      

\begin{sloppypar}
The crucial result in the derivation of the asymptotic thermal state \eqref{therm} is thus the expression 
\eqref{UtUt'} which states that the dynamical correlations 
$\langle \psi | U_{ij} (t)  U_{kl} (t') | \psi \rangle$ where $| \psi \rangle$ is a typical pure bath state 
of macroscopically well-defined energy $E$, are identical to the respective canonical correlation 
functions with the temperature determined by $E$ via \eqref{T}. This result stems 
from $\langle U_{ij} U_{kl}(\tau) U_{lk} (\tau) U_{ji}  \rangle_E \ll \exp(N)$ for $N \gg 1$
and from the equality \eqref{UtauU} of microcanonical and canonical correlation functions. 
These correlation functions are a priori related by 
\begin{equation}
\langle U_{ij} U_{kl}  (\tau) \rangle_T = \frac{1}{Z}
\int dE n(E) e^{-E/T} \langle U_{ij} U_{kl}  (\tau) \rangle_E  
\end{equation}       
for any environment. Due to the Boltzmann's relation \eqref{n}, the function $n(E)\exp(-E/T)/Z$ is 
essentially a Gaussian of variance $\propto N$ peaked at the energy $E$ 
related to $T$ by \eqref{T}. For the environment considered in this paper, 
the microcanonical correlations $\langle U_{ij} U_{kl}  (\tau) \rangle_E$ 
vary on macroscopic energyscales and hence are equal to canonical averages as 
given by \eqref{UtauU}. 
\end{sloppypar}

To better comprehend how the canonical distribution for $S$ emerges, we consider 
the microcanonical form of the rate \eqref{Gammacan}, i.e., with 
$Z^{-1}\sum_\alpha \exp(-E_\alpha/T)$ replaced by $D^{-1} \sum_{|\alpha \rangle \in {\cal H}_E}$. 
We write this expression as 
\begin{equation}
\Gamma_{ik} (E)=\frac{1}{D} \sum_{|\alpha \rangle \in {\cal H}_E} \gamma_{ik}^{(\alpha)}
\end{equation} 
where the $E$-dependence of the rate is noted explicitly and the definition 
$\gamma_{ik}^{(\alpha)}=\sum_{\beta} \delta(E_\beta-E_\alpha+\omega_{ik}) 
|\langle \beta |U_{ik}| \alpha \rangle|^2$ is used.
With these notations and the property $U^\dag_{ki}=U^{\phantom{\dag}}_{ik}$, we find 
\begin{equation}
\Gamma_{ki} (E)=\frac{1}{D} \sum_{|\alpha \rangle \in {\cal H}_{E+\omega_{ik}}} \gamma_{ik}^{(\alpha)} 
= \frac{n(E+\omega_{ik})}{n(E)} \Gamma_{ik} (E+\omega_{ik}) .
\end{equation}  
The last equality follows from $D=n(E)\delta E$. It clearly 
shows that the detailed balance relation for the rates $\Gamma_{ik}$ 
results from the Boltzmann's relation \eqref{n} and from the fact that $\Gamma_{ik}$, 
which is the Fourier transform of $\langle U_{ik} U_{ki}  (\tau) \rangle_E$ with frequency $\omega_{ik}$, 
is practically the same for the macroscopic energies $E$ and $E+\omega_{ik}$. 

We finally comment on the weak coupling regime considered in the previous section. 
In this regime, the self-energies describing the influence of the bath on $S$ 
are well approximated by their expansions to second order in the system-bath interaction 
Hamiltonian $H_I$. The detailed balance relation is satisfied by 
the rates \eqref{Gammacan} deduced from the second-order self-energies. 
If we suppose, for example, that a part of the Hamiltonian 
$H_S$ describing the intrinsic dynamics of $S$ is of the same order \cite{Capek}, 
contributions stemming from this perturbative self-Hamiltonian term 
must be added to the second-order self-energies. Consequently, the zeroth-order asymptotic state 
is no longer given by a master equation with rates satisfying the detailed balance relation. 
Our results thus indicate, as the previous derivations \cite{Diu,Tasaki,Mahler1,CT}, that 
the canonical distribution is an attribute of the weak coupling regime. 
 
 \section{Conclusion}
 \label{sec:conclusion}

In this paper, we have studied the time evolution of a system $S$ under the 
influence of a heat bath of harmonic oscillators. The reduced dynamics of $S$ 
has been derived exactly from the Schr\"odinger 
equation of the isolated composite system consisting of $S$ and its bosonic environment. 
Contrary to the usual assumption of a bath initially at thermal equilibrium, 
a {\it pure} initial bath state was considered. We have shown that, for almost all 
initial bath states of macroscopically well-defined energy, the system $S$ 
relaxes into canonical equilibrium, provided the coupling to the bath is weak enough. 
The relaxation of $S$ into a canonical mixed state has thus been obtained without 
performing any statistical averaging. In other words, 
no equilibrium state is assumed a priori in our derivation. 
The bath though initially in a pure state induces the thermalisation of the system $S$. 
The temperature $T$ of the asymptotic thermal state of $S$ is determined by the macroscopic 
energy of the initial bath state. It equals the microcanonical 
temperature, defined from the bath density of states, corresponding to this energy. 
We have also found that, for the overwhelming majority of initial bath states 
of macroscopically well-defined energy, the time evolution of $S$ cannot be distinguished 
from the one obtained assuming that the boson bath is initially at thermal 
equilibrium at temperature $T$.

We emphasize that though these results have been derived for a boson bath, a substantial part 
of our approach is applicable to other environments. As discussed in the previous section, 
the thermalisation of the system $S$ results essentially from two characteristics 
of the bath: (i) the bath density of states verifies the Boltzmann's relation 
(ii) the microcanonical correlation functions of the bath operators coupled to $S$ 
vary on macroscopic energyscales. 
While the first feature is of great generality for physical environments, 
it is not so obvious for the second one. 
It would then be interesting to study the microcanonical correlations of 
the coupling operators of other ``realistic'' environments. 
Another interesting generalisation of the present work would be to add the possibility 
for the system $S$ of interest to exchange also particles with its environment and to investigate whether, 
under these conditions, a reservoir initially in 
a ``typical'' pure state induces the relaxation of $S$ into grand-canonical equilibrium.   

\begin{acknowledgement}

We thank R. Chitra for a careful reading of the manuscript.

\end{acknowledgement}

\end{document}